\documentclass[preprint,showpacs,english,superscriptaddress]{revtex4-1}
\usepackage{amsmath}
\usepackage{graphicx}
\usepackage{dcolumn}
\usepackage{bm}
\usepackage{subdepth}
\usepackage{multirow}
\usepackage{pbox}

\begin{document}

\title{Temperature Dependence of Angular Momentum Transport Across Interfaces}
\author{Kai Chen}
\affiliation{Department of Physics, University of Arizona, Tucson, Arizona 85721}
\author{Weiwei Lin}
\affiliation{Department of Physics and Astronomy, Johns Hopkins University, Baltimore, Maryland 21218}
\author{C. L. Chien}
\affiliation{Department of Physics and Astronomy, Johns Hopkins University, Baltimore, Maryland 21218}
\author{Shufeng Zhang}
\affiliation{Department of Physics, University of Arizona, Tucson, Arizona 85721}

\begin{abstract}
Angular momentum transport in magnetic multilayered structures plays a central role in spintronic physics and devices. The angular momentum currents or spin currents are carried by either quasi-particles such as electrons and magnons, or by macroscopic order parameters such as local magnetization of ferromagnets. Based on the generic interface exchange interaction, we develop a microscopic theory that describes interfacial spin conductance for various interfaces among non-magnetic metals, ferromagnetic and antiferromagnetic insulators. Spin conductance and its temperature dependence are obtained for different spin batteries including spin pumping, temperature gradient and spin Hall effect. As an application of our theory, we calculate the spin current in a trilayer made of a ferromagnetic insulator, an antiferromagnetic insulator and a non-magnetic heavy metal. The calculated results on the temperature dependence of spin conductance quantitatively agree with the existing experiments.

\end{abstract}

\date{\today }
\maketitle

\section{introduction}

In spintronics, one of the most important variables is spin current which describes the propagation of angular momentum information through magnetic and non-magnetic media \cite{Maekawa}. There are a number of different carriers that contribute to spin current. In non-magnetic metals, the carriers are conduction electrons while for magnetic insulators, the angular momentum carriers are magnons or spin waves. When these different carriers meet at interfaces, they transfer the angular momentum via interfacial exchange interaction. For example, the spin pumping describes a precessing ferromagnet transferring its long wavelength magnon current to an electron spin current in the adjacent metallic layer \cite{Tserkovnyak, Maekawa13}, and the spin Seebeck effect addresses the spatially non-uniform thermal magnon diffusion \cite{Uchida08, Uchida08b, Jaworski, Giles}.

Recent experiments have shown that angular momentum current transfer at interfaces is a general phenomenon for many combinations of materials as long as the low-energy carriers (quasi-particles or order parameters) of the materials have nonzero angular momentum \cite{Kajiwara, Cornlissen, Goennenwein, Shi16, Han16,Yang14, Saitoh15, Chien, Hahn, Frangou}. In a trilayer made of a ferromagnetic insulator (FI) layer (YIG) sandwiched between two non-magnetic metallic layers (Pt), it has been observed that a charge current applied in one of the metal layers can result in a charge current in the other layer via magnon-mediated spin current propagation \cite{Kajiwara, Cornlissen, Goennenwein, Shi16, Han16}. The observed signal is much more profound at high temperature, indicating that a simple model based on a temperature independent interfacial mixing conductance would fail to describe the experimental findings \cite{Goennenwein, Han16}. Other recent experiments demonstrated that
the spin current can flow from a ferromagnetic insulator to a non-magnetic metallic (NM) layer with a thin antiferromagnetic insulator (AFI) in between \cite{Yang14, Saitoh15, Chien, Hahn}. Furthermore, the spin propagation efficiency
is much enhanced at high temperature when compared with the device without the AFI layer \cite{Chien, Yang14}. These findings call for a more comprehensive theoretical model which is capable of addressing the angular momentum current across interfaces between different materials at finite temperature.

There are a number of existing theoretical models for the spin conductance (SC) near interfaces. In spin pumping, the SC or mixing conductance between a ferromagnetic layer and non-magnetic metallic layer has been calculated at zero temperature using first principle methods \cite{Xiake}. In spin Seebeck effect, the SC between the FI and NM layers has been studied by model Hamiltonians and the resulting SC is highly temperature dependent \cite{Zhang12b}. Thus, the spin conductances for the thermally driven spin Seebeck effect and for the spin pumping are quite different even though the interface is identically same. There are also theoretical studies involving AFI layer. Ohnuma \emph{et al.} calculated the spin current due to a temperature difference across the AFI and NM interface \cite{Maekawa14}. Cheng \emph{et al.} studied spin pumping from an AFI layer to a NM layer \cite{Niu14}. Recently, Rezende \emph{et al.} introduced a mixing conductance for the interfaces between FI and AFI layers phenomenologically without calculating its temperature or material dependence \cite{Rezende16b}.

In this paper,  we develop a theory to formulate the SC for interfaces with different material combinations by using a generic interface exchange Hamiltonian,  with an emphasis on the temperature dependence of the SC. The paper is organized as follows. In Sec.~II, we introduce the concept of spin battery for three different spin current generators, define the interface SC, and summarize our results for various interface SCs in Table I. In Sec.~III, we provide detailed models and calculations to support the results in Table I. In Sec. IV, we apply the above SCs to the FI/AFI/NM trilayers and compare our results to available experiments. The excellent agreement with the experimental data are obtained. We conclude the paper in Sec. V.

\section{Summary of interface spin conductance}

The interface SC is defined as the ratio of the spin or angular momentum current across the interface to the spin voltage drop at the two sides of the interface. The spin voltage is provided by a spin battery. Followed the three spin current generators introduced in Ref. \cite{Bauer13}, we define the spin battery voltage in each case before calculating the SC.

First, the spin voltage of the ``spin pumping battery'' \cite{Brataas02},  which is generated by an external microwave source such that the magnetization of ferromagnetic layer undergoes precession motion in the ferromagnetic resonance (FMR) condition, can be defined as
\begin{equation}
\mathbf{ V}_{\rm sp} = \frac{\hbar}{2} \mathbf{ m}\times \frac{d\mathbf{ m}}{dt}, 
\end{equation}
where $\mathbf{ m}$ is the dimensionless unit vector representing the direction of the magnetization of the layer. It is understood that the spin pumping battery provides non-equilibrium magnons with zero wave number ($k=0$). 

The second spin battery is created by a temperature gradient across a FI layer \cite{Zhang12, Rezende14}. The presence of the position-dependent temperature $T=T(x)$ in the FI layer ($x<0$) leads to a non-uniform local magnon density
\[
n(x)=\int d\varepsilon_{\mathbf{q}} g_m^{\rm F}(\varepsilon_{\mathbf{q}})N_0(\varepsilon_{\mathbf{q}},T)
\]
where $N_0(\varepsilon_{\mathbf{q}}, T)= [e^{\varepsilon_{\mathbf{k}}/k_BT}-1]^{-1} $ is the Bose-Einstein distribution function and $g_m^{\rm F}(\varepsilon_{\mathbf{q}})$ is the FI magnon density of states. The magnon diffusion generates a magnon current in the FI layer. When the magnon current flows to the interface, a non-equilibrium magnon density is accumulated 
near the interface. These non-equilibrium magnon accumulation becomes a spin voltage that can excite spin degree of freedom at the other side of  the interface. In the open circuit condition (i.e., an isolated FI layer without a contacting layer), the magnon accumulation is proportional to the magnon diffusion length. Thus, we define the thermally driven spin battery voltage as
\begin{equation}
\mathbf{ V}_{\rm th} = \lambda_{\rm F} \frac{d (k_BT)} {dx} \mathbf{ m}
\end{equation}
where $k_B$ is the Boltzmann constant and $\lambda_{\rm F}$ is the magnon diffusion length within the FI layer.

The third battery is built up in a non-magnetic layer such as Pt with a large spin Hall angle. When an in-plane current is applied to the NM layer, a spin Hall current flowing perpendicular to the charge current is generated. Similar to the magnon accumulation for magnetic materials, electron spin accumulation is built near the interface and scales with the spin diffusion length in the open circuit condition \cite{Zhang00}. The spin Hall battery voltage in this case is
\begin{equation}
\mathbf{ V}_{\rm sh} = e \theta_{\rm sh} \rho \lambda_{\rm N} \hat{ \mathbf{z}} \times \mathbf{j}_e
\end{equation}
where $e$ is the electron charge, $\theta_{\rm sh}$ is the spin Hall angle, $ \lambda_{\rm N}$ is the spin diffusion length within the NM material,
$\rho$ is the resistivity, $\hat{\mathbf{ z}}$ is the unit vector 
normal to the interface, and $\mathbf{ j}_e$ is the electron current density.

We emphasize a few points on the above definitions: 1) we have chosen the unit of the spin battery to be that of energy, 2) the spin battery is a vector which characterizes the direction of the angular momentum (note that the spin pumping battery is transverse to $\mathbf{ m}$ and the temperature gradient battery is parallel to $\mathbf{ m}$), 3) the battery ``stores'' different forms of spin angular momenta: zero-wave number magnons for spin pumping battery, magnon accumulation with a broad distribution of wave numbers for the thermal battery,  and electron spin accumulation for the spin Hall battery. 

These spin batteries, in Eqs.~(1), (2), and (3), are defined for an isolated layer, i.e, in the absence of spin current. When the battery is connected to a layer which is capable of carrying spin momenta, a spin current flows in the neighboring layer as well as in the battery layer. Thus, both internal spin current (within the battery layer) and external spin current will ``consume" spin angular momentum. However, the comparison between charge and spin batteries on the internal and external resistance or conductance shows one fundamental difference: the electric current is conserved but the spin current is not, thus the addition of the resistance in series is no longer valid for the spin resistance \cite{Datta}; we shall illustrate in later sections on how to calculate the spin current with many layers or many spin conductors in series. The main goal of the present paper is to calculate the SC at finite temperatures, for interfaces between different materials and for three different batteries. We shall first tabulate our calculated results in Table I. The explanation of the Table I is given below and the detailed derivation of these results will be given in the next Section.

\begin{table}[tbp]
\small
\newcolumntype{U}{>{\centering\arraybackslash}p{3.6cm}}
\newcolumntype{V}{>{\centering\arraybackslash}p{2.5 cm}}
\newcolumntype{W}{>{\centering\arraybackslash}p{4.5cm}}
\renewcommand\thetable{I}
\caption{List of spin conductance $G_{\rm int} a^2$ ($a$ is the lattice constant) of several magnetic interfaces driven by different batteries. In these bilayer structures, the spin current across the interface is $\mathbf{ j}_{s}=G_{\rm int}\mathbf{V} /[2\pi(1+\epsilon)]$ where $\epsilon$ characterizes a backflow spin current and will be calculated in late sections. The Table gives the dependence of the SC on temperature $T$, interface coupling strength $J_{\rm int}$, electron density of states at Fermi level $g_e(E_F)$, Curie temperature $T_C$, and N\'eel temperature $T_N$.}
\centering 
\renewcommand{\arraystretch}{1.2} 
\begin{tabular}{V|U | U| W}
\hline\hline 
Batteries & \small Interface   		&  $H_{\rm int}$                	&      $G_{\rm int} a^2$          \\
\hline 
\multirow{2}{*}{\pbox{20 cm}{Spin \\ pumping}} 	&FI/NM        			& $ J_{\rm int} a_{0}^{+}c_{\mathbf{k}\uparrow}^{+}c_{\mathbf{k}\prime\downarrow}$
& $ \Big( J_{\rm int}g_{e}(E_F) \Big)^2$      											 \\
\cline{2-4}
&FI/AFI    			& $ J_{\rm int}a_{\mathbf{q}_1}^{+} a_{\mathbf{q}_2} a_{0} \beta_{\mathbf{q}_3}$
&$\tfrac{J_{\rm int}^2}{
(k_B T_{C})(k_B T_{\rm N})}\left(\frac{T}{T_C}\right)^{2} \frac{T}{T_{\rm N}} $                        \\ 

\hline  
\multirow{2}{*}{\pbox{20 cm}{Temperature\\ gradient}} 	&FI/NM         			& $ J_{\rm int}a_{\mathbf{q}}^{+}c_{\mathbf{k}\uparrow}^{+}c_{\mathbf{k}\prime\downarrow}$
& $  \Big( J_{\rm int}g_{e}(E_F) \Big)^2\left(\frac{T}{T_C}\right)^{{3/2}}$           \\
\cline{2-4}
&FI/AFI       			& $ J_{\rm int}a_{\mathbf{q}}^{+}\alpha_{\mathbf{q}\prime}$ &     $ \tfrac{J_{\rm int}^2}{
(k_B T_{C})(k_B T_{\rm N})}\left(\frac{T}{T_C}\right)^{{1/2}}\frac{T}{T_{\rm N}}$     \\ 

\hline 
\multirow{3}{*}{Spin Hall} 		&NM/FI ($\bm{\mu}_{s}\perp \mathbf{m}$)         & $ J_{\rm int}a_{0}^{+}c_{\mathbf{k}\uparrow}^{+}c_{\mathbf{k}\prime\downarrow}$
& $ \Big( J_{\rm int}g_{e}(E_F) \Big)^2$          \\
\cline{2-4}
&NM/FI ($\bm{\mu}_{s}\parallel \mathbf{m}$)     & $ J_{\rm int}a_{\mathbf{q}}^{+}c_{\mathbf{k}\uparrow}^{+}c_{\mathbf{k}\prime\downarrow}$
& $\Big( J_{\rm int}g_{e}(E_F) \Big)^2\left(\frac{T}{T_C}\right)^{3/2}$           \\ 
\cline{2-4}
&NM/AFI ($\bm{\mu}_{s}\parallel \mathbf{n}$)    & $ J_{\rm int}\left(\alpha_{\mathbf{q}}^{+}+\beta_{\mathbf{q}}\right)c_{\mathbf{k}\uparrow}^{+}c_{\mathbf{k}\prime\downarrow}$
& $ \Big( J_{\rm int}g_{e}(E_F) \Big)^2 \left(\frac{T}{T_{\rm N}}\right)^{2}$           \\
\hline\hline 
\end{tabular}
\end{table}

Table I shows the spin conductance for three spin batteries. In the first two cases in which the battery layer is a FI, we consider two bilayers, FI/NM and FI/AFI. In the third case, the battery is the NM layer and we consider NM/FI and NM/AFI interfaces. In all bilayers, the total spin current also depends on the backflow \cite{Tserkovnyakb}: when the battery generates a spin current in the neighboring layer, a spin or magnon accumulation will be established in the layer, which in turn, flows a portion of the spin current back to the battery, resulting a smaller interface spin current. The backflow parameter, $\epsilon$, is determined by the ratio of the spin conductance at the interface to that in the layers. In a bilayer structure, the backflow parameter for three batteries has the same form, $\epsilon=G_{\rm int}/G_{\rm L}+G_{\rm int}/G_{\rm R}$ where $G_{\rm L/R}$ is the spin conductance of the left/right layer, see next section for details.

For spin pumping at FI/NM interfaces, the angular momentum current 
conversion occurs between the zero wave number magnons in the FI layer and the conduction electron spins in the NM layer \cite{Maekawa13}. The spin conductance in this case has been identified as the mixing conductance. The temperature dependence is unimportant since the conduction electron distribution is weakly dependent on temperature. For other interfaces, i.e., FI/AFI, the spin conductance involves conversion from FI magnons to AFI magnons with broadly distributed wave numbers. Since the density of the magnons is highly temperature dependent, one expects a similar dependence for the SC. The SC in Table I is for low temperatures  (lower than N\'eel or Curie temperatures) where the temperature dependence can be analytically derived. For higher temperatures, analytical expressions are unavailable; we will present the numerical results in later Sections. The spin conductances for the temperature gradient battery are shown with the same two interfaces, FI/NM and FI/AFI. In both cases, there are strong temperature dependence. 

Spin conductance for the spin Hall battery is also summarized. It is interesting to note that the electron spin current from the spin Hall battery can excite two types of magnons: coherent zero wave number magnons which represent the uniform magnetization precession or spin transfer torque ($\boldsymbol{\mu}_s \perp \mathbf{m}$), and incoherent magnons that produce a dc magnon current ($\boldsymbol{\mu}_s \parallel \mathbf{m}$ or $\boldsymbol{\mu}_s \parallel \mathbf{n}$). When driven by the spin Hall effect, the interface spin conductance is either same as the spin pumping conductance or the thermal conductance depending on the relative direction between the electron spin accumulation and the magnetization. We will further discuss these in next section. It is noted that the magnetic metal is not included in this paper because of an additional complication: a magnetic metal has both magnons and conduction electron spins, and thus spin current in different layers will involve much more channels; we will leave such complication for further studies.

\section{Calculation of spin conductance}

In this Section, we derive the conductances shown in Table I. We start with specifying the model Hamiltonian for each layer and determining the dispersion relations of equilibrium quasi-particles including the ferromagnetic and antiferromagnetic magnons. By using a generic exchange coupling between spins of the two layers at the interface, we compute the interface spin conductance and total spin current include backflow. 

\subsection{The model systems and spin Hamiltonians}

We first consider simple models for each individual layer. For nonmagnetic metals, the spin current carriers are conduction electrons whose dispersion relations are described by free electron model, i.e., $\varepsilon_{k} = (\hbar k)^2/2m_e$. For FI or AFI, we model the spin Hamiltonian below,
\begin{equation}
H=\pm J_{ex}\sum_{\langle i, j \rangle} \mathbf{S}_{i}\cdot\mathbf{S}_{j}
 - H_{\rm ext}\sum_{i}S_{i}^{z} -K\sum_{i}\left( S_{i}^{z} \right)^{2}
\end{equation}
where $J_{ex}$ is the exchange constant between nearest neighbors, $H_{\rm ext}$ is the external magnetic field applied in the $z$ direction and $K$ is the easy axis anisotropy constant.

When choosing the minus sign in the above Hamiltonian, the spin lattice has a ferromagnetic ground state. Within the spin wave approximation, one can readily
obtain the low-energy quasiparticle spectrum as
\begin{equation}
H_{\rm F}=\sum_{\mathbf{q}} \varepsilon_{\mathbf{q}}^{\rm F}a_{\mathbf{q}}^{+}a_{\mathbf{q}}
\end{equation}
where $\varepsilon_{\mathbf{q}}^{\rm F}=2J_{ex}SZ(1-\gamma_{\mathbf{q}})+2KS+\gamma_{0}H_{\rm ext}$ is the magnon dispersion, $Z$ is the number of nearest neighbors, $S$ the magnitude of each atomic spin and $\gamma_{\mathbf{q}}=1/Z\sum_{\bm{\delta}}e^{i\mathbf{q}\cdot \bm{\delta}}$ where $\bm{\delta}$ runs over all nearest neighbor positions. $\Delta_{\rm F}=2KS$ is the FI magnon gap.

With the positive sign,  the Hamiltonian describes an antiferromagnetic lattice. Within the spin wave approximation, the magnon spectra are
\begin{equation}
H_{\rm A}=\sum_{\mathbf{q}}\left( \varepsilon_{\mathbf{q}}^{\alpha}\alpha_{\mathbf{q}}^{+}\alpha_{\mathbf{q}} +\varepsilon_{\mathbf{q}}^{\beta}\beta_{\mathbf{q}}^{+}\beta_{\mathbf{q}} \right)
\end{equation}
where $\alpha_{\mathbf{q}}$ and $\beta_{\mathbf{q}}$ represent two branches of magnon and $\varepsilon_{\mathbf{q}}^{\alpha,\beta}=J_{ex} SZ\sqrt{(1+2K/JZ)^2-\gamma_{\mathbf{q}}}~\pm\gamma_{0}H_{\rm ext}$. $\Delta _{\rm A}=\sqrt{2KS\times J_{ex}SZ}$ is the AFI magnon gap . 

\begin{figure}[tbp]
\includegraphics[width=12cm]{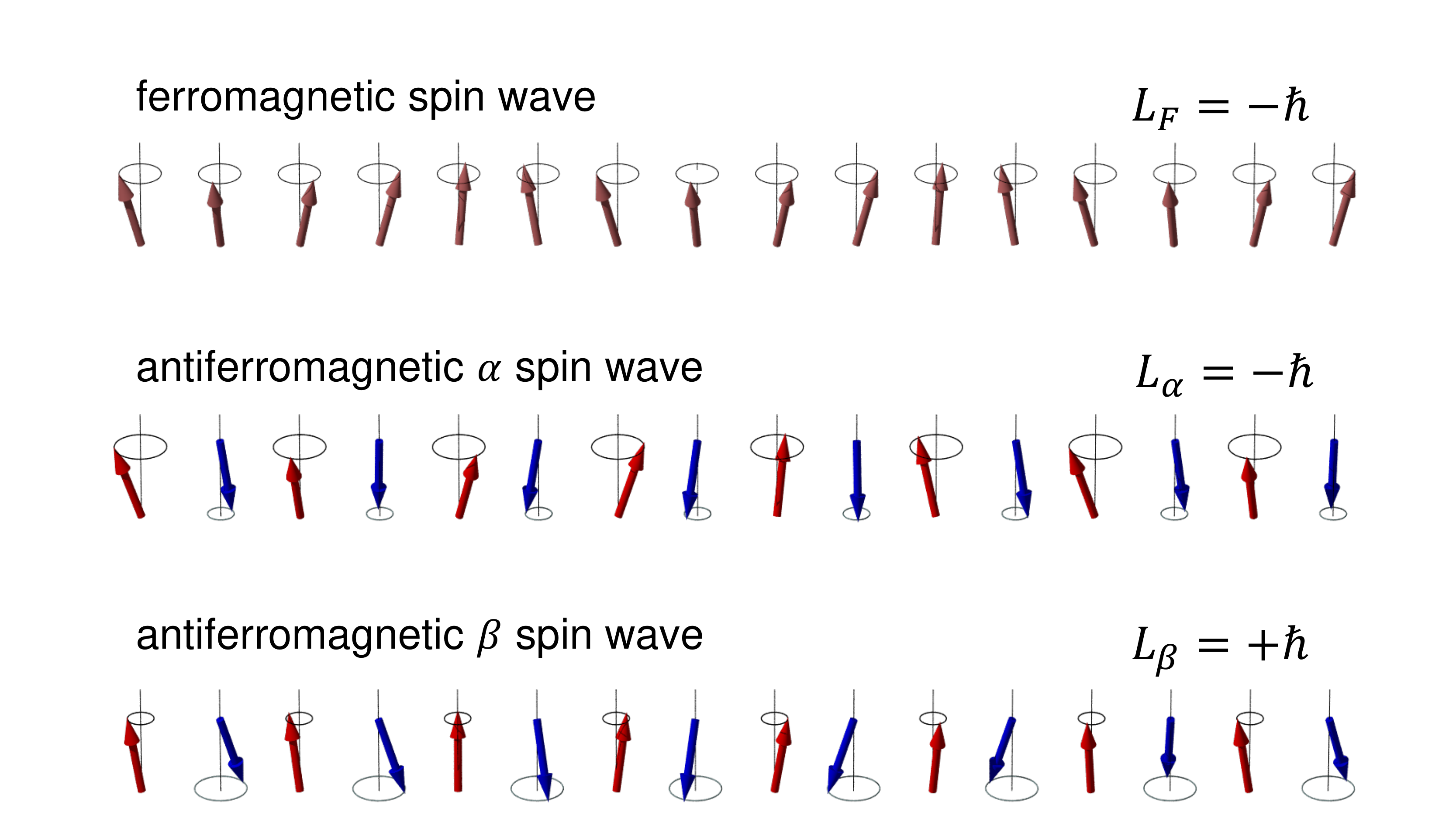}
\caption{(Color Online) Spin waves in ferromagnets and antiferromagnets. The brown, red and blue arrows are the spins on the FI lattice, AFI sublattice $A$ and AFI sublattice $B$, respectively.}
\end{figure}

There are two important distinctions between the FI and AFI magnons. First, the FI magnon has a small energy gap determined by the anisotropy while the AFI magnon has a much larger gap because it scales with the geometrical average of the exchange constant and the anisotropy. Another distinction is that each F magnon carries an angular momentum $-\hbar$ with respect to the
magnetization direction while in the AF lattice, a magnon in one branch ($\alpha_\mathbf{ q}$) carries $-\hbar$ and the other ($\beta_\mathbf{ q}$) carries $\hbar$. In Fig. 1, we depict spin configuration of a FI magnon and a AFI magnon in each of the two branches. A $\alpha_{\mathbf{q}}$ magnon represents the mode with a larger precession angle for sublattice $A$ (Red) than $B$ (Blue), i.e., $\theta_A > \theta_B$. While both $\theta_A$ and $\theta_B$ depend on $\mathbf{ q}$, the
angular momentum is $L_{\alpha}= -N_{\rm A} S \hbar \left[(\theta_{\mathbf{q}}^A)^2-(\theta_{\mathbf{q}}^B)^2\right] \equiv - \hbar$ for a $\alpha$ magnon
and $L_{\beta}= \hbar$ for a $\beta$ magnon, where $N_{\rm A}$ is the number of spins in the AFI lattice. In the absence of the external magnetic field, $\alpha_{\mathbf{q}}$ and $\beta_{\mathbf{q}}$ magnons have exactly same energy, indicating that these two degenerate magnon branches are equally populated at any temperature, and thus there is no net magnetization or spin current at equilibrium.

Having specified the angular momentum carriers in each layer, we now introduce the spin interaction between two materials in contact. A generic exchange interaction at the interface between two spins would be simplest and universal,
\begin{equation}
H_{\rm int} = -J_{\rm int} \sum_{i} \mathbf{ S}_i^{(L)} \cdot \mathbf{ S}_i^{(R)}
\end{equation}
where $\mathbf{S}_i^{(L)}$ ($\mathbf{ S}_i^{(R)}$) represents the spin at the interface of the left (right) layer. For  the FI or AFI layers, $\mathbf{S}_i$ refers to the spin at the local site, while for the NM, $\mathbf{S}_i$ denotes the spin of conduction electrons at the interface.

\subsection{Spin conductance of a spin pumping battery}

The spin pumping battery has widely been used for the generation of the spin current in NM layers.
The SC has first been formulated via  interfacial reflection
and transmission coefficients in the scattering approach \cite{Tserkovnyak}. Other models \cite{Maekawa13}, including a simple linear response theory \cite{CZ15}, yield essentially same result.  Here we briefly re-derive it with Eq.~(7) for the FI/NM interface and then continue
with the derivation for the FI/AFI interface.

The second quantization of Eq. (7) at the FI/NM interface is
\begin{equation}
H_{\rm int} =-J_{\rm int}\sqrt{2S_{\rm F}}\sum_{\mathbf{kk'} \mathbf{q}} \left(a_{\mathbf{q}}^{+}c_{\mathbf{k}\uparrow}^{+}c_{\mathbf{k'}\downarrow}+H.c. \right)\delta_{\mathbf{k',k+q}}
\end{equation}
where $c_{\mathbf{k}\sigma}^{+}~(c_{\mathbf{k}\sigma})$ is the conduction electron creation (annihilation) operator, $N_{\rm F}~(N_{\rm N})$ is number of lattice sites of FI (NM) at the interface and $S_{\rm F}$ is the magnitude of each FI spin. The spin current across the interface is,
\begin{equation}
j_{s}=\left\langle\frac{1}{i A_I}\left[ \sum_{\mathbf{q}}a_{\mathbf{q}}^{+}a_{\mathbf{q}},
H_{\rm int}\right]\right \rangle
\end{equation}
where $[,]$ is the quantum commutator, $\langle \rangle$  refers to the average over all states and $A_I$ is the interface cross area. Use the rough interface approximation, we don't impose the momentum conservation in Eq. (8). By placing Eq.~(8) into Eq.~(9) and by utilizing the random phase approximation, we find
\begin{equation}
j_{s}=\frac{2\pi J_{\rm int}^2S_{\rm F}}{ N_{\rm F}N_{\rm N} A_I}\sum_{\mathbf{k} \mathbf{k'}\mathbf{q}} \left[ (N_{\mathbf{q}}^{\rm F}+1)(1-f_{\mathbf{k}\uparrow})f_{\mathbf{k'}\downarrow} -
 N_{\mathbf{q}}^{\rm F}f_{\mathbf{k}\uparrow}(1-f_{\mathbf{k'}\downarrow}) \right]\delta(\varepsilon_{\mathbf{q}}^{\rm F}+\varepsilon_{\mathbf{k}}-\varepsilon_{\mathbf{k}'})
\end{equation}
where $N_{\mathbf{q}}^{\rm F}$ and $f_{\mathbf{k'}s}$ are the magnon and electron distribution functions. In thermal equilibrium, the magnons and electrons can be described by the Boson and Fermion statistics.

For the spin pumping voltage, the magnon distribution is the sum of the thermal magnon $N_0(\varepsilon_{\mathbf{q}},T)$ and coherent $q=0$ magnons $\delta_{\mathbf{q}0}N_{\rm F}S_{\rm F}\sin^{2}\theta$  representing the uniform precession driven by microwave magnetic field, where $\theta$ is the magnetization precession angle. The energy of a $q=0$ magnon is given by the FMR frequency $\omega$, i.e., $\varepsilon_{q=0}^{\rm F}=\hbar \omega$. Inserting the distribution function into Eq.~(10), we find
\begin{equation}
j_{s}^{\rm sp,NM}=2\pi \hbar J_{\rm int}^2 S_{\rm F}^2 a_{\rm N}^4 g_{e}^2(E_{\rm F})\omega \sin^{2}  \theta
\end{equation}
where $a_{\rm N}$ is the lattice constant of the NM material and $g_{e}(E_{\rm F})$ the electron density of states near Fermi energy. Under the FMR condition, we identify $\omega \sin^2\theta$ as the dc component of $\mathbf{m}\times \frac{d \mathbf{m}}{dt}$. Compare with the definition of the spin conductance
$\mathbf{ j}_s = G_{\rm F/N}^{\rm sp} \mathbf{ V}_{\rm sp}/2\pi$, we find,
\begin{equation}
G_{\rm F/N}^{\rm sp}=8\pi^2 J_{\rm int}^2 S_{\rm F}^2 a_{\rm N}^4 g_{e}^2(E_{\rm F}).
\end{equation}
The above SC, after discarding the unimportant constants, is listed in the first row of Table I. We note that Ohnuma \emph{et al.} have already derived the SC using similar method, but expressed the result in terms of  ferromagnetic susceptibility \cite{Maekawa13}. By replacing the susceptibility with the Lindhard susceptibility of a non-magnetic metal, one will directly get the mixing conductance derived here.

Next, we calculate the spin pumping conductance for a FI/AFI interface. The second quantization of $H_{\rm int}$ in Eq.~(7) gives the coupling between FI and AFI magnons. The lowest order terms refer to two magnon interactions. The angular momentum conservation limits the possible two-magnons processes to $a_{0}(\alpha_{\mathbf{q'}}^+ +\beta_\mathbf{ q'})$ and its complex conjugate. However, such process is prohibited by the energy conservation: the energy of the FMR frequency or $q=0$ magnon is too small to excite any magnon in the AFI. Thus, the angular momentum current across the interface must go through at least four magnon processes. By expanding Eq.~(7) to four magnon operators, we obtain a number of terms which satisfy both energy and angular momentum conservation. For example, the term $ a_{\mathbf{ q}_1}^+ a_0a_{\mathbf{ q}_2}\beta_{\mathbf{q}_3}$ represents the transfer of the angular momentum in the FI by annihilating a $q=0$ and two thermal magnon of the FI layer, and simultaneously annihilating a $\beta$ magnon in the AFI layer, as long as $ \varepsilon_{\mathbf{q}_1}^{\rm F}=\varepsilon_{0}^{\rm F}+\varepsilon_{\mathbf{q}_2}^{\rm F}+\varepsilon_{\mathbf{q}_3}^{\alpha}$. After tedious but straightforward calculations, we find the spin current across the interface via such four magnon processes can be written as
\begin{eqnarray}
j_s^{\rm sp, AFI}=\frac{\pi J_{\rm int}^2 S_{\rm A}}{8 N_{\rm F}N_{\rm A}}  \sum_{\mathbf{q}_1\mathbf{q}_2\mathbf{q}_3} &
   \left( \zeta_{\mathbf{q}_3}^2 +  \zeta_{\mathbf{q}_3}^{-2}\right)  &
 \delta \left (\varepsilon_{\mathbf{q}_1}^{\rm F}-\varepsilon_{\mathbf{q}_2}^{\rm F}-\varepsilon_{\mathbf{q}_3}^{\alpha}-\varepsilon_{0}^{\rm F}\right) \\
&& \left[ (N_{\mathbf{q}_1}^{\rm F}+1)  N_{\mathbf{q}_2}^{\rm F}N_{\mathbf{q}_3}^{\alpha}N_{q=0}^{\rm F}-N_{\mathbf{q}_1}^{\rm F}(N_{\mathbf{q}_2}^{\rm F}+1)(N_{\mathbf{q}_3}^{\alpha}+1)(N_{q=0}^{\rm F}+1)\right] \notag
\end{eqnarray}
where $\zeta_{\mathbf{q}}^2=\lvert (\theta_{A}-\theta_B)/(\theta_A + \theta_B)\rvert$ \cite{umklapp} and $\theta_A~(\theta_B)$ is the precession angle for a given magnon defined in Fig. 1,  $N_{\mathbf{q}_{1/2}}^{\rm F}$ and $N_{q=0}^{\rm F}$ are the FI magnon distribution functions, and $N_{\mathbf{q}_3}^{\alpha}$ are the distribution functions of AFI $\alpha$ magnons; in the long wavelength limit, $\zeta_{\mathbf{q}}^2 \simeq \varepsilon_{\mathbf{q}}/J_{ex}S_{\rm A}Z$. 
By inserting the ferromagnetic resonance driven magnon distribution function, $N_{\mathbf{q}}=N_0(\varepsilon_{\mathbf{q}},T)+\delta_{\mathbf{q}0}N_{\rm F}S_{\rm F}\sin^{2}\theta$, we find the SC at FI/AFI interface due to the $ a_{\mathbf{ q}_1}^+ a_0a_{\mathbf{ q}_2}\beta_{\mathbf{q}_3}$ process, $G_{\rm A/F}^{\rm sp} =2\pi \mathbf{j}_s^{\rm sp, AFI}/\mathbf{V}_{\rm sp}$, is
\begin{eqnarray}
G_{\rm A/F}^{\rm sp}=a_{\rm F}^5 a_{\rm A}^2 \frac{J_{\rm int}^2 S_{\rm A}}{32 k_B T} \int d\varepsilon_{\mathbf{q}} \int &d\varepsilon_{\mathbf{q}^\prime} &
 \left( \zeta_{\mathbf{q}^\prime}^2 +  \zeta_{\mathbf{q}^\prime}^{-2}\right) g_m^F(\varepsilon_{\mathbf{q}}) g_m^{\rm A}(\varepsilon_{\mathbf{q}^\prime}) g_m^F(\varepsilon_{\mathbf{q}}+\varepsilon_{\mathbf{q}^\prime}) \\
&&{\rm csch}^{2}\frac{\varepsilon_{\mathbf{q}}}{2k_B T} {\rm csch}^{2}\frac{\varepsilon_{\mathbf{q}^\prime}}{2k_B T}
 {\rm csch}^{2}\frac{\varepsilon_{\mathbf{q}}+\varepsilon_{\mathbf{q}^\prime}}{2k_B T} \notag
\end{eqnarray}
where $a_{\rm F}~(a_{\rm A})$ is the FI (AFI) lattice constant and $g_{m}^{\rm A/F}(\varepsilon)$ is the AFI/FI magnon density of states. For temperatures much lower than the Curie and N\'eel temperatures, Eq.~(14) reduces to the value listed in the second row of Table I where the unimportant numerical factors are discarded. The term $a_{\mathbf{ q}_1}^+ a_0a_{\mathbf{ q}_2}\alpha_{\mathbf{q}_3}^+$ makes identical contribution to the spin conductance that shown in Eq. (14). Notice that the interaction in Eq. (7) also contains other four magnon terms involving three AFI magnons and one $q=0$ FI magnon like $a_{0}\alpha_{\mathbf{q}_1}^+\alpha_{\mathbf{q}_2}^+\alpha_{\mathbf{q}_3}$ and so on. Below the N\'eel temperature, the spin pumping conductance from those terms can be estimated as $a_{\rm F}^{-1}a_{\rm A}^{-1}\frac{J_{\rm int}^2}{k_{B}^2 T_C T_N} \left( \frac{T}{T_N}\right)^5$.  The total spin pumping conductance is the sum of all these contributions.

As we have discussed earlier, the total spin current depends on the backflow. The backflow can be easily included if the layer thickness is much larger than the relevant length scales such as the spin or magnon diffusion lengths. The spin current provided by the spin battery decays in the layer; this creates a spin accumulation or magnon accumulation that drive a backflow spin current. One may introduce a spin conductance $G_{\rm N}=(h/2e^2)(1/\rho \lambda_{\rm N})$ as the spin conductance for the NM layer and similarly, $G_{\rm F}$ and $G_{\rm A}$ for the FI and AFI layers. The Onsager reciprocal relation can be used to determine the backflow current \cite{Brataas12} such that the total spin current across the interface is reduced by $(1+ \epsilon )^{-1}$ where the backflow factor $\epsilon = G^{\rm sp}_{\rm int} (G_{\rm F}^{-1} + G_{\rm N}^{-1})$. We will discuss the relative magnitudes of these SCs when we apply our theory to a concrete multilayer.

\subsection{Spin conductance of a temperature gradient spin battery}

 The spin Seebeck current across a FI/NM bilayer has been theoretically studied based on the diffusion theory of thermal magnons \cite{Zhang12, Rezende14}. Far from the interface, the temperature gradient perpendicular to the interface drives a magnon current. The magnon current leads to a non-equilibrium magnon accumulation near the interface. In contrary to the spin pumping case where the non-equilibrium magnons only exists for $q=0$, there is a broad magnon spectrum distribution. For the FI/NM interface, the interaction in the spin wave approximation is same as Eq.~(8) and the expression of Eq.~(10) remains valid. However, we need to replace the magnon distribution by,
\begin{equation}
N_\mathbf{ q}^{\rm F} = \frac{1}{e^{ (E_\mathbf{ q}-\mu_m(x))/k_BT} -1}
\end{equation}
where we have introduced 
the spatial dependent magnon chemical potential, $\mu_m(x)$. At equilibrium, $\mu_m(x)$ is identically zero. In the presence of magnon accumulation, $\mu_m (x)$ characterizes the number of the non-equilibrium magnons,
\begin{equation}
\delta n(x) \simeq g_m^{\rm F}(T) \mu_m(x) .
\end{equation}
where $g_m^{\rm F}(T)=-\int d \varepsilon g_m^{\rm F}(\varepsilon)\partial _{\varepsilon} N_0(\varepsilon, T) $. By inserting the non-equilibrium distribution functions, $N_{\mathbf{q}}^{\rm F}$ and $f_{\mathbf{ k} \sigma}=f_0-\frac{\partial f_0}{\partial E_\mathbf{ k}}\mu_{\sigma}(0^+)$ into Eq. (10), we find the spin current at the interface is,
\begin{equation}
j_{s}^{\rm th,NM}(0)=\frac{G_{\rm F/N}^{\rm th}}{2\pi} \left[ \mu_m(0^-) - \mu_s(0^+)\right]
\end{equation}
where $\mu_s(0^+)=\mu_{\uparrow}(0^+)-\mu_{\downarrow}(0^+)$ is the spin split chemical potential at the interface and
\begin{equation}
G_{\rm F/N}^{\rm th}=\frac{\pi^2 J_{\rm int}^2 S_{\rm F}}{k_B T}a_{\rm N}^3 a_{\rm F}^4 g_{e}(E_{\rm F})^2 \int d\varepsilon_{\mathbf{q}} g_m^{\rm F}(\varepsilon_{\mathbf{q}}) \varepsilon _{\mathbf{q}} {\rm csch}^2 \frac{\varepsilon_{\mathbf{q}} }{2k_B T}
\end{equation}
is the thermal driven interface spin conductance. If the temperature is lower than the Curie temperature of the FI, the SC reduces to
a simple $T^{3/2}$ power law listed in Table I. The inclusion of the backflow can be similarly done; the calculated backflow parameter $\epsilon$ has same forms as that of the spin pumping, with one distinction: in the present case, $G_{\rm F}$ is the spin
conductance for the longitudinal spin current (proportional to the magnon-diffusion length), while $G_{\rm F}$ in the spin pumping
battery is for the transverse spin current where the spin dephasing length is much smaller. 

The second interface for the thermally driven spin battery is  the FI/AFI interface in which the thermal magnons in the FI transfer to the magnons in the AFI layer. In contrast to the spin pumping battery where the two magnon process is prohibited, the thermal magnons have a broad spectrum of the magnon energy in the FI layer and thus it is possible to directly transfer one FI magnon to one AFI magnon, i.e., the interface spin exchange interaction in the form of $J_{\rm int} a_\mathbf{ q} \alpha_{\mathbf{ q}'}^+$ leads to a spin current across the interface,
\begin{equation}
j_{s}^{\rm th, AFI}=\frac{2\pi J_{\rm int}^2 S_{\rm F} S_{\rm A}}{A_I }\sum_{\mathbf{qq'}} \left( \zeta_{\mathbf{q'}}^2 +  \zeta_{\mathbf{q'}}^{-2}\right) \left[ N_{\mathbf{q}}^{\rm F}(N_{\mathbf{q'}}^{\alpha}+1)-(N_{\mathbf{q}}^{\rm F}+1)N_{\mathbf{q}'}^{\alpha}\right] \delta(\varepsilon_{\mathbf{q}}^{\rm F} - \varepsilon_{\mathbf{q}'}^{\alpha})
\end{equation}
where $N_{\mathbf{q}}^{\rm F}$ and $N_{\mathbf{q}'}^{\alpha}$ are the FI and AFI magnon distribution functions respectively. Notice that only the transmission from FI magnon to the $\alpha_q$ branch of AFI magnon can conserve energy and angular momentum at the same time. Following the similar procedure in deriving the SC of FI/NM spin interface, we find the interface current
\begin{equation}
j_{s}^{\rm th, AFI}(0)=\frac{G_{\rm F/A} ^{\rm th}}{2\pi}\left[\mu_m(0^-)-\mu_m(0^+)\right]
\end{equation}
where $\mu_{m}(0^{-/+})$ measures the non-equilibrium FI/AFI magnon accumulation at the interface, and the interface conductance is
\begin{equation}
G_{\rm F/A}^{\rm th}=\frac{\pi^2 J_{\rm int}^2 S_{\rm F} S_{\rm A}}{ k_B T} a_{\rm F}^2 a_{\rm A}^2 \int d\varepsilon_{\mathbf{q}}   \left( \zeta_{\mathbf{q}}^2 +  \zeta_{\mathbf{q}}^{-2}\right)
g_m^{\rm F}(\varepsilon_{\mathbf{q}} ) g_m^{\rm A}(\varepsilon_{\mathbf{q}} )  {\rm csch}^2 \frac{\varepsilon _{\mathbf{q}}}{2k_B T}
\end{equation}

\subsection{Spin conductance with the spin Hall battery}

The sources of the spin current in previous two batteries reside in the FI layer. We next consider a non-magnetic layer with a large spin Hall angle as a spin battery. As we have introduced earlier, an in-plane charge current creates a spin voltage in the direction of $\hat{\mathbf{ z}}\times \mathbf{ j}_e $ due to the spin Hall effect. For a ferromagnetic layer in contact with the spin Hall battery, the spin current would depend on the relative direction between the magnetization $\mathbf{ m}$ and the spin voltage. If  $\mathbf{ m} $ is perpendicular to the spin voltage $\hat{\mathbf{ z}}\times \mathbf{ j}_e $, the spin current entering the ferromagnetic layer decays within very small length, resulting a spin torque at the interface. This spin conductance at the FI/NM interface is the same as the mixing conductance defined in Eq. (12). There are quite extensive studies on the magnetization switching by the spin Hall current \cite{Liu12, Liu12b}. In the case where $\mathbf{ m} \parallel \hat{\mathbf{ z}}\times \mathbf{ j}_e $, the spin Hall battery creates non-equilibrium magnons in the FI layer. The spin conductance for the parallel case is identical to the $G^{\rm th}_{\rm F/N}$ shown in Eq. (18). Both spin conductances have been already calculated previously \cite{Brataas, Zhang12b},  we have listed them in Table I. Here we present the calculation for the NM/AFI interfaces.

The Hamiltonian in Eq. (7) within the spin wave approximation is 
\begin{align}
H_{\rm int} = -J_{\rm int} \sqrt{\frac{2S_{\rm A}}{N_{\rm N} N_{\rm A}}} \sum_\mathbf{ kk'q}\left[ \zeta_{\mathbf{q}} (\alpha_{\mathbf{q}}^+ + \beta_{\mathbf{q}})c_\mathbf{ k \downarrow}^+ c_\mathbf{ k', \uparrow} + H.c.\right] \delta_{\mathbf{k',k+q}} \notag \\
+\left[\zeta_{\mathbf{q}}^{-1}(\alpha_{\mathbf{q}}^+ - \beta_{\mathbf{q}})c_\mathbf{ k \downarrow}^+ c_\mathbf{ k', \uparrow}  + H.c.\right]\delta_{\mathbf{k',k+q+G}}
\end{align}
where the first term is normal scattering, the second term stands for the Umklapp scattering \cite{Niu14} and $\mathbf{G}$ is half of the reciprocal NM lattice vector. Again, we don't impose the momentum conservation at the interface in the following calculation.
The angular momentum current across the interface is
\begin{align}
j_{s}=\frac{2\pi J_{\rm int}^{2}S_{\rm A}}{ N_{\rm A} N_{\rm N} A_I}& \sum_{\mathbf{kk^{\prime }q}}\left( \zeta_{\mathbf{q}}^2 +  \zeta_{\mathbf{q}}^{-2}\right)
\left[(N_{\mathbf{q}}^{\alpha }+1)(1-f_{\mathbf{k}\uparrow})f_{\mathbf{k}^{\prime }\downarrow }-
N_{\mathbf{q}}^{\alpha }f_{\mathbf{k}\uparrow}\left( 1-f_{\mathbf{k}^{\prime }\downarrow }\right)   \right]\delta
\left( \varepsilon _{\mathbf{k}}+\varepsilon _{\mathbf{q}}^{\alpha }-\varepsilon_{\mathbf{k}^{\prime }}\right) \notag\\
 - &\left( \zeta_{\mathbf{q}}^2 +  \zeta_{\mathbf{q}}^{-2}\right)\left[(N_{\mathbf{q}}^{\beta }+1)f_{\mathbf{k}\uparrow }(1- f_{\mathbf{k}^{\prime }\downarrow
})-N_{\mathbf{q}}^{\beta }\left( 1-f_{\mathbf{k}\uparrow }\right) f_{\mathbf{k}^{\prime }\downarrow
}  \right]\delta \left(\varepsilon _{\mathbf{k}}-\varepsilon _{\mathbf{q}}^{\beta
}-\varepsilon _{\mathbf{k}^{\prime }}\right)
\end{align}%
By placing the non-equilibrium distribution of the battery into Eq.~(23), we find,
\begin{equation}
G_{\rm N/A}^{\rm th} = \frac{2\pi^2 J_{\rm int}^{2} S_{\rm A}}{k_{B}T}g_{e}^{2}(E_{\rm F}) a_{\rm N}^{4} a_{a}^{3}\int d\varepsilon_{\mathbf{q}} \left( \zeta_{\mathbf{q}}^2 +  \zeta_{\mathbf{q}}^{-2}\right) g_{m}^{\rm A}(\varepsilon_{\mathbf{q}})
\varepsilon_{\mathbf{q}}\text{csch}^{2}\left( \frac{\varepsilon_{\mathbf{q}}}{2k_{B}T}\right).
\end{equation}
The above SC is applied to the case when the spin battery is parallel to the staggered magnetization of the AFI. The superscript ``th'' (thermal) indicates the above spin conductance involves the spin convertance between conduction electrons and magnons across the whole spectrum instead of only the $k=0$ mode. When they are perpendicular, a spin current driven spin torque on the AFI has been proposed; this will involve the coherent AFI magnon
generation by the spin Hall battery \cite{Niu14}.

\section{Application for multilayered structures at high temperatures}

In Table I, we have listed the interface spin current and conductance of bilayers with semi-infinite thickness for each layer. Experimentally, there can be more than two layers whose thicknesses are comparable to the spin or magnon decaying length. Furthermore, experiments are usually carried out at room temperature which is not much lower than the Curie or N\'eel temperatures. For example, the spin current with a thin NiO is largest near the N\'eel temperature \cite{Chien, Yang14, Saitoh15} . Thus, in the following, we describe how the interface SCs in Table I are applied to multilayers with finite thickness and how these SCs changes at temperatures near or above critical temperatures.

\subsection{Boundary conditions for spin currents in multilayers}

Similar to the electron spin transport in metallic multilayers, we need boundary conditions and the spin/magnon
diffusion equations within each layer. The SC in Table I will be used as boundary conditions at $x=0$,
\begin{equation}
j_s (0^+) = j_s (0^-) = \frac{G_{\rm int}}{2\pi} \left[ \mu (0^+)-\mu (0^-) \right]
\end{equation}
where $G_{\rm int}$ is the interface SC for a particular interface, and $\mu (0^+) $ [$\mu (0^-)$] represents the chemical potential of the electrons or magnons at the right [left] interface. Within each layer, including the battery layer, the spin current is given by
\begin{equation}
j_s (x) = j_b (x) - \sigma \frac{d \mu (x)}{dx}
\end{equation}
where $j_b (x)$ is the source spin current in the battery layer and is zero elsewhere. To illustrate how these boundary conditions along with the diffusion equations determine the spin current in the entire multilayers, we take an example of a trilayer consisting of
FI/AFI/NM, driven by a temperature gradient battery across the FI layer.  The spin/magnon chemical potentials in each layer has the following forms: $\mu_m = C_1 \exp(x/\lambda_{\rm F})$ in the FI layer $(x<0)$, $\mu_m (x) = C_2 \exp(-x/\lambda_{\rm A})
+ C_3 \exp(x/\lambda_{\rm A})$ in the AFI layer ($0<x<d_{\rm A}$) and $\mu_s (x) =C_3
\exp(-x/\lambda_{\rm N})$  in the NM layer ($x>d_{\rm A}$) where $d_{\rm A}$ is the thickness of the AFI layer, $\lambda_{\rm F}$, $\lambda_{\rm A}$, and $\lambda_{\rm N}$ are the diffusion lengths in each layers. of the AFI layer. By using the boundary conditions, Eq.~(25) for the interfaces FI/AFI and AFI/NM at $x=0$ and $x=d_{\rm A}$, four constants of integration $C_i$ ($i=1-4$) are readily determined. While the expression of the spin current is rather lengthy and cumbersome for an arbitrary thickness of the AFI, it takes a particularly simple form if we assume 1) the thickness of the AFI is much smaller than $\lambda_{\rm A}$ so that there is no spin current decay in the AFI layer, and 2) the interface spin conductance of the FI/AFI is much larger than that of the NM/AFI interface. We find the spin current in the NM layer is
\begin{equation}
\mathbf{j}_{s}^{tri} (x) =\frac{G_{\rm N/A}^{\rm th}\exp(-x/\lambda_{\rm N})}{1+G_{\rm N/A}^{\rm th}/G_{\rm F} +G_{\rm N/A}^{\rm th}/G_{\rm N}} \frac{ \mathbf{V}_{\rm th}}{2\pi}
\end{equation}%
If we further approximate the FI as a good spin sink  so that one may neglect the second term in the denominator \cite{Rezende14}. Comparing the spin current above to that of the bilayer FI/NM, i.e., without the AFI insertion, we have
\begin{equation}
\eta_{\rm th} \equiv \frac{j_{s}^{tri}}{j_s^{bi}} =1+\frac{ (a-1) G_{\rm N}}{G_{\rm N/A}^{\rm th}+G_{\rm N}}
\end{equation}
where
\begin{equation}
a= \frac{G_{\rm N/A}^{\rm th}}{G_{\rm F/N}^{\rm th}} =C'\left( \frac{J_{\rm NiO/Pt}}{J_{\rm YIG/Pt}} \right)^{2}\left( \frac{T}{T_{\rm N}} \right)^2
\left(\frac{T}{T_c} \right)^{-3/2}
\end{equation}
$C'$ is a numerical constant of the order of 1, $J_{\rm NiO/Pt}$ and $J_{\rm YIG/Pt}$ are the interface exchange constants and we have used the interface SC of Table I.

Interestingly, if $a\gg 1$, i.e., the spin conductance for NiO-Pt interface is much larger than YIG-Pt interface, the enhancement with
an AFI layer insertion is significant and the largest occurs at high temperatures. We will further address the enhancement in the next subsection. Next, we consider the same trilayer structure by replacing the thermal battery with a spin pumping battery. Within the same approximation, the spin current in the NM layer is
\begin{equation}
\mathbf{ j}_{s}^{\rm sp, tri} (x) =\frac{\hbar}{4\pi}\frac{G_{\rm F/A}^{\rm sp} \exp(-x/\lambda_{\rm N}) }{1+G_{\rm F/A}^{\rm sp}/G_{\rm N/A}^{\rm th}+ G_{\rm F/A}^{\rm sp}/G_{\rm N}} \mathbf{ m}\times \frac{d \mathbf{ m}}{dt}.
\end{equation}
Notice that at the YIG/NiO interface, the battery is magnetization precession in the YIG layer. Thus, we use $G^{\rm sp}_{\rm F/A}$ as the interface conductance. At the NiO/Pt interface, the spin battery is the magnon accumulation with broad wave number distribution, and the interface spin conductance is given by $G^{\rm th}_{\rm N/A}$. Again, the spin pumping current vanishes at low temperature, reflecting the fact that magnon or spin current is blocked by either the FI/AFI or AFI/NM interface at low temperatures. The spin current enhancement with the AFI layer is,
\begin{equation}
\eta_{\rm sp}  =1+\frac{ (b-1) (G_{\rm N}+G_{\rm F/N}^{\rm sp})}{ G_{\rm F/A}^{\rm sp} \left(1+G_{\rm N}/G_{\rm N/A}^{\rm th}\right)+G_{\rm N}}
\end{equation}
where $ b= G_{\rm F/A}^{\rm sp}/G_{\rm F/N}^{\rm sp} $.

\subsection{Modeling spin current at elevated temperatures}

As our theory is built on the spin wave approximation, one would expect the theory not applicable to high temperatures, in particular, near the transition temperature. However, the most interesting features with the AFI layers discovered experimentally occur at a temperature near or even above the N\'eel temperature \cite{Saitoh15, Chien}. Thus, it is desirable to extend the formalism with reasonably approximations.

\begin{figure}[tbp]
\includegraphics[width=15cm]{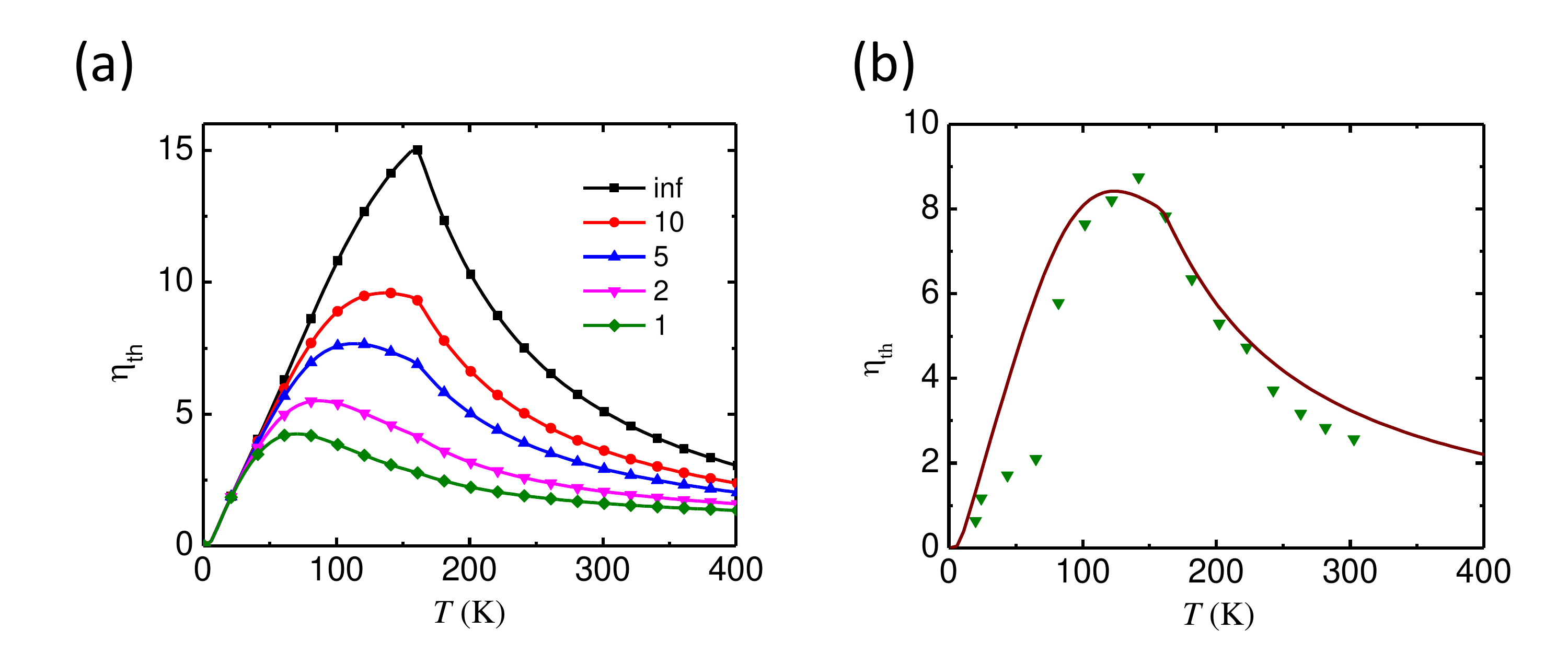}
\caption{(Color Online) The spin Seebeck signal enhancement factor, $\eta_{\rm th}$, as a function of temperature for various spin conductance of the NM layer from $G_{\rm N} = 1, 2, 5, 10, \infty $ $ (10^{18} {\rm m}^{-2})$, calculated by using Eq. (28). The parameters are $a_{\rm F}=1.39$ nm, $a_{\rm A}=0.42$ nm, $T_C=560$ K, $T_N=160$ K ($t_{NiO}=0.6$ nm), $g_e(E_F)=3n_e/2E_F$ with $n_e=5\times 10^{22} {\rm cm}^{-3}$ and $E_F=5$ eV. $J_{\rm YIG/Pt} = $ 0.07 eV and $J_{\rm NiO/Pt}=0.13$ eV are the $sd$ constant at the interface.  (a) The enhancement factor for a number of spin conductance of the NM layer. (b) Comparison of the experimental points \cite{Chien} with the theoretical curve for $G_{\rm N}=6.7\times 10^{18} {\rm m}^{-2}$.
}
\end{figure}

The spin transport near transition temperatures is in general an unresolved theoretical issue. While there are a number of approximate methods to treat the critical phenomena, no rigorous theory exists for a wide range of temperatures. Here we should remain to use the spin wave approximation with one limitation: above the transition temperature, the spin wave approximation breaks down since spin correlation length becomes finite. In early theories and neutron scattering experiments, it was indeed found that the spin wave with long wavelengths loses its meaning, but the short wavelength magnon remains intact \cite{Guggenheim, Lindgard}. For example, the spin correlation length of NiO is 
\begin{equation}
\xi = l\left( \frac{T-T_{\rm N}}{T_{\rm N}}\right)^{-\nu}
\end{equation}
where $l=1.2a_{\rm NiO}$, $a_{\rm NiO}=0.42$ nm is the lattice constant of NiO and $\nu=0.64$ \cite{Lindgard}. The magnon whose wavelength is shorter than $\xi$  has well-defined dispersion relation \cite{Guggenheim}, indicating the presence of short-range AF spin correlations. We thus modify our spin wave approximation by assuming a cutoff energy $\hbar \omega_{q_c}$ where $q_c = 1/\xi$ such that $N_q =0$ for $q<q_c$. When the temperature increases, the long wavelength magnons do not participate transport. With this modification, we are able to address the spin current propagation for a wide range of temperatures.

As an example, we consider the same FI/AFI/NM (YIG/NiO/Pt) trilayer. At high temperatures, we no longer use Table I for the interface SC. Instead, we will use the general expression, Eqs. (18) and (24) by placing the cutoff energy as a lower bound of the integration. In Fig.2(a), we show the spin current enhancement as a function of the temperature for a thermal battery for different NM spin conductance. As the temperature increases, the number of magnons and the interface conductance increase, thus the spin current, mediated by the magnons in the AFI layer increases. When the temperature reaches to the N\'eel temperature of  the AFI (Note that the Curie temperature of YIG is much higher), the number of magnons participating the angular momentum transport begins to decrease due to removing of the long wavelength magnons. Meanwhile, the spin current in the bilayer structure keeps increasing with temperature when $T_{\rm N}<T<T_{\rm C}$. Thus, both the spin conductance $G_{\rm N/A}^{\rm th}$ and the enhancement factor $\eta_{\rm th}$ are maximum near the N\'eel temperature. We notice that spin current peak at the transition temperature has been obtained by Okamoto by using a different approach \cite{Okamoto}.  The enhancement is reduced as the NM layer SC decreases due to enhanced back flow, consistent with Eq. (29). Interestingly, the peak position occurs at a lower temperature for smaller SC of the NM layer; this can be explained as follows. When $G_{\rm N/A}^{\rm th}$ becomes larger than $G_{\rm N}$, the spin current in the YIG/NiO/Pt trilayer saturates, while for the YIG/Pt bilayer, spin current continues to increase with temperature since $G_{\rm N/F}^{\rm th}$ remains smaller than $G_{\rm N}$. Notice that the calculated $\eta_{\rm th}$ deviates the $T^{1/2}$ law even at low temperatures due to the large AF magnon gap. In Fig.~2(b), we compare our calculations with $G_{\rm N}$ measured in previous publication \cite{Ding12}; the agreement is considered to be excellent \cite{Chien}.

\section{Discussions and conclusions}

We have developed a theory based on spin current transfer at interfaces. The different spin current carriers are mutually converting
via an interfacial spin exchange Hamiltonian. Within the spin wave approximation, we are able to explicitly formulate the SC for different sources of the spin current (spin batteries) and for different interfaces at finite temperature. 

We point out that the SC studied here is for quasi-particle spin transport, i.e., the spin current carriers are incoherent low-energy quasiparticles, which is different from the ``super-current'' carried by the macroscopic classical magnetization (coherent magnons), or the order parameter. For the quasiparticle transport, the quantum statistics governs the temperature dependent properties. In general, both incoherent and coherent magnons contribute to the spin transport.

Our theory is particularly effective to be used for multilayered structure at finite temperature with arbitrary layer thickness. Using the diffusion equation for each layer along with the interface SCs, one is able to determine the spatial and temperature dependence of the spin current. The spin battery, which is an extension of the spin pumping battery introduced earlier \cite{Brataas02}, is a convenient concept that can be used to describe the spin current flow. In analogy with an electric battery: the spin battery has just one terminal while the electrical battery must have at least two terminals because of the conservation law imposed to the charge current. For the spin battery, one can still use spin Ohm's law, i.e., 
$dV_s(x)/dx = j_s (x) G_s^{-1} (x)$ where $G_s^{-1}(x)$ is a local spin resistivity. Due to non-conservative nature of the spin current,
the spin current $j_s (x)$ is no longer a constant throughout the layers. Thus, the spin Ohm's law alone (even if $G_s$ is known) cannot determine the spin current. In this paper, we have provided a general scheme for computing the spin current.

Our theory provides a natural explanation to the temperature dependence of current propagation through FI and AFI insulators. Recent experiments on Pt/YIG/Pt have confirmed our earlier prediction \cite{Zhang12b}. The spin current enhancement by inserting a thin NiO layer at the interface of the YIG/NM \cite{Saitoh15, Chien}, quantitatively supports our theory. The other theories based on the order parameter spin transport \cite{Takei,Slavin} have not taken into account finite temperature effects.

K. C. and S. Z. were supported by National Science Foundation under Grant number ECCS-1404542. The work at JHU was supported as part of SHINES, an EFRC funded by the US DOE Basic Energy Science under award number SC0012670.


\begin{thebibliography}{99}
\bibitem{Maekawa} \textit{Spin Current}, edited by S. Maekawa, S. O. Valenzuela, E. Saitoh,
and T. Kimura (Oxford University, 2012).

\bibitem{Tserkovnyak} Y. Tserkovnyak, A. Brataas, and G. E. W. Bauer, Phys. Rev. Lett. \textbf{88}, 117601 (2002).

\bibitem{Maekawa13} Y. Ohnuma, H. Adachi, E. Saitoh, and S. Maekawa, Phys. Rev. B \textbf{87}, 014423 (2013).

\bibitem{Uchida08} K. Uchida, H. Adachi, T. Ota, H. Nakayama, S. Maekawa, and E. Saitoh, Appl. Phys. Lett. \textbf{97}, 172505 (2010).

\bibitem{Uchida08b} K. Uchida, S. Takahashi, K. Harii, J. Ieda, W. Koshibae, K. Ando, S. Maekawa, and E. Saitoh, Nature \textbf{455}, 778 (2008).

\bibitem{Jaworski} C. M. Jaworski, J. Yang, S. Mack, D. D. Awschalom, J. P. Heremans, and R. C. Myers, Nat. Mater. \textbf{9}, 898 (2010).

\bibitem{Giles} B. L. Giles, Z. Yang, J. S. Jamison, and R. C. Myers, Phys. Rev. B \textbf{92}, 224415 (2015).

\bibitem{Kajiwara} Y. Kajiwara \emph{et al}., Nature (London) \textbf{464}, 262 (2010).

\bibitem{Cornlissen} L. J. Cornelissen, J. Liu, R. A. Duine, J. Ben Youssef, and B. J. van Wees, Nat. Phys. \textbf{11}, 1022 (2015).

\bibitem{Goennenwein} S. T. B. Goennenwein, R. Schlitz, M. Pernpeintner, K. Ganzhorn, M. Althammer,
R. Gross, and Hans Huebl, Appl. Phys. Lett. \textbf{107}, 172405 (2015).

\bibitem{Shi16} J. Li, Y. Xu, M. Aldosary, C. Tang, Z. Lin, S. Zhang, R. Lake, and J. Shi,
Nat. Commun. \textbf{7}, 10858 (2016);

\bibitem{Han16} H. Wu, C. H. Wan, X. Zhang, Z. H. Yuan, Q. T. Zhang, J. Y. Qin, H. X. Wei, X. F. Han, and S. Zhang, Phys. Rev. B \textbf{93},060403 (2016).

\bibitem{Saitoh15} Z. Qiu, J. Li, D. Hou, E. Arenholz, A. T. NDiaye, A. Tan, K. Uchida, K. Sato, Y. Tserkovnyak, Z. Q. Qiu, and E. Saitoh, arXiv:1505.03926.

\bibitem{Chien} W. Lin, K. Chen, S. Zhang, C. -L. Chien, arXiv:1603.00931.

\bibitem{Yang14} H. Wang, C. Du, P. C. Hammel, and F. Yang, Phys. Rev. Lett. \textbf{113}, 097202 (2014).

\bibitem{Hahn} C. Hahn, G. de Loubens, V. V. Naletov, J. Ben Youssef, O. Klein, and Michel Viret, Europhys. Lett. \textbf{108}, 57005 (2015).

\bibitem{Frangou} L. Frangou, S. Oyarzun, S. Auffret, L. Vila, S. Gambarelli, and V. Baltz, Phys. Rev. Lett. \textbf{116}, 077203 (2016).

\bibitem{Xiake} K. Xia, P. J. Kelly, G. E.W. Bauer, A. Brataas, and I. Turek, Phys. Rev. B \textbf{65}, 220401 (2002); X.-T. Jia, K. Liu, K. Xia, and G. E.W. Bauer, Europhys. Lett. \textbf{96}, 17005 (2011).

\bibitem{Zhang12b} Steven S.-L. Zhang and S. Zhang, Phys. Rev. B \textbf{86}, 214424 (2012).

\bibitem{Maekawa14} Y. Ohnuma, H. Adachi, E. Saitoh, and S. Maekawa, Phys. Rev. B \textbf{89}, 174417 (2014).

\bibitem{Niu14} R. Cheng, J. Xiao, Q. Niu, and A. Brataas, Phys. Rev. Lett. \textbf{113}, 057601 (2014).

\bibitem{Rezende16b} S. M. Rezende, R. L. Rodriguez-Suarez, and A. Azevedo, Phys. Rev. B \textbf{93}, 054412 (2016).

\bibitem{Bauer13} M. Weiler \textit{et al.}, Phys. Rev. Lett. \textbf{111}, 176601 (2013).

\bibitem{Brataas02} A. Brataas, Y. Tserkovnyak, G. E. W. Bauer, and B. I. Halperin, Phys. Rev. B \textbf{66}, 060404(R) (2002).

\bibitem{Zhang12} Steven S.-L. Zhang and S. Zhang, Phys. Rev. Lett. \textbf{109}, 096603 (2012).

\bibitem{Rezende14} S. M. Rezende, R. L. Rodriguez-Suarez, R. O. Cunha, A. R. Rodrigues, F. L. A. Machado, G. A. Fonseca Guerra, J. C. Lopez Ortiz, and A. Azevedo, Phys. Rev. B 89, 014416 (2014).

\bibitem{Zhang00} S. Zhang, Phys. Rev. Lett. \textbf{85}, 393 (2000).

\bibitem{Datta} K. Y. Camsari, S. Ganguly, and S. Datta, Sci. Rep. \textbf{5} 10571 (2015).

\bibitem{Tserkovnyakb} Y. Tserkovnyak, A. Brataas, and G. E. W. Bauer, Phys. Rev. B \textbf{66}, 224403 (2002).

\bibitem{CZ15} K. Chen and S. Zhang, Phys. Rev. Lett. \textbf{114}, 126602 (2015).

\bibitem{umklapp} The $\zeta_{\mathbf{q}}^{-2}$ term in Eq. (13) emerges from the Umklapp scattering at the FI/AFI interfaces. Eq. (22) also includes such scattering for a NM/AFI interface, which has been pointed out in \cite{Niu14}.

\bibitem{Brataas12} A. Brataas, Y. Tserkovnyak, G. E. W. Bauer, and P. J. Kelly, in \textit{Spin Current}, edited by S. Maekawa, S. O. Valenzuela, E. Saitoh, and T. Kimura (Oxford University, 2012).

\bibitem{Liu12} L. Liu, O. J. Lee, T. J. Gudmundsen, D. C. Ralph, and R. A. Buhrman, Phys. Rev. Lett. \textbf{109}, 096602 (2012).

\bibitem{Liu12b} L. Liu, C. -F. Pai, Y. Li, H. W. Tseng, D. C. Ralph, and R. A. Buhrman, Science \textbf{336}, 555 (2012).

\bibitem{Brataas} A. Brataas, Y. V. Nazarov, and G. E.W. Bauer, Eur. Phys. J. B \textbf{22}, 99-110 (2001).

\bibitem{Guggenheim} J. Skalyo, Jr., G. Shirane, R. J. Birgeneau and H. J. Guggenheim,
Phys. Rev. Lett. \textbf{23}, 1394 (1969); R. J. Birgeneau, H. J. Guggenheim, and G. Shirane,
Phys. Rev. B \textbf{8}, 304 (1973).

\bibitem{Lindgard} T. Chatterji, G. J. McIntyre, and P.-A. Lindgard, Phys. Rev. B \textbf{79}, 172403 (2009).

\bibitem{Takei} S. Takei, T. Moriyama, Teruo Ono, and Y. Tserkovnyak, Phys. Rev. B \textbf{92}, 020409(R) (2015).

\bibitem{Slavin} R. Khymyn, I. Lisenkov, V. S. Tiberkevich, A. N. Slavin, and B. A. Ivanov, arXiv:1511.05785.

\bibitem{Okamoto} Satoshi Okamoto, Phys. Rev. B \textbf{93}, 064421 (2016).

\bibitem{Ding12} Z. Feng, J. Hu, L. Sun, B. You, D. Wu, J. Du, W. Zhang, A.
Hu, Y. Yang, D. M. Tang, B. S. Zhang, and H. F. Ding, Phys. Rev. B \textbf{85%
}, 214423 (2012).

\end{thebibliography}
\end{document}